\documentstyle [12pt]{article}
\begin{document}
\baselineskip 20.0 pt
\par
\mbox{}
\vskip -1.25in\par
\mbox{}
 \begin{flushright}
\makebox[1.5in][1]{ITP-SB-92-58}\\
\makebox[1.5in][1]{CTP-2156}\\
\makebox[1.5in][1]{October 1992}
 \end{flushright}
 \vskip 0.25in

\begin{center}
{\bf On the KP Hierarchy, $\hat{W}_{\infty}$ Algebra,
and Conformal SL(2,R)/U(1) Model}\\
{\bf --- The Classical and Quantum Cases}{\footnote{Invited talk given
by the second author at the XXI International Conference on Differential
Geometric Methods in Theoretical Physics, Nankai Institute of Mathematics,
Tianjin, China; June 5-9, 1992; to appear in Proceedings.}} \\
\vspace{40 pt}
{Feng Yu$^{\dagger}$ and Yong-Shi Wu$^{\dagger\dagger}$}\\
\vspace{20 pt}
{{\it $\dagger$ Institute for Theoretical Physics, State
University of New York}}\\
{{\it Stony Brook, New York 11794, U.S.A.}}\\
\vspace{20 pt}
{{\it $\dagger\dagger$Center for Theoretical Physics, MIT}}\\
{{\it Cambridge, MA 02139, U.S.A.}}\\
{{\it and}}\\
{{\it Department of Physics, University of Utah}}\\
{{\it Salt Lake City, Utah 84112, U.S.A.}}\\

\vspace{40pt}
{\large ABSTRACT}\\
\end{center}
\vspace{10 pt}

We give a unified description of our recent results on the the
inter-relationship between the integrable infinite KP
hierarchy, nonlinear $\hat{W}_{\infty}$ current algebra and conformal
noncompact $SL(2,R)/U(1)$ coset model both at the classical and
quantum levels. In particular, we present the construction of a quantum
version of the KP hierarchy by deforming the second KP Hamiltonian
structure through quantizing the $SL(2,R)_k/U(1)$ model and
constructing an infinite set of commuting quantum $\hat{W}_{\infty}$
charges (at least at $k$=1).

\newpage

\noindent {\it Introduction and Summary}

One of the most exciting recent developments in mathematical physics
is the revelation of the close relationship between infinite
dimensional (extended) conformal algebras, nonlinear integrable
differential systems and 2d exactly solvable field theories, including
conformal field models, 2d quantum gravity and string theory.
A well-known example is the Virasoro algebra, which is known to be
isomorphic to the second Hamiltonian structure of the KdV hierarchy,
and appears as the operator product algebra in 2d minimal conformal models,
and as constraint algebra in $c<1$ string theory. Similar relations happen
to Zamolochikov's extended $W_N$ algebras ($N\geq 3$), extension of
the Virasoro algebra generated by currents of higher spin ($2\leq s\leq N$).

In our recent study$^{1-4}$, we generalize this profound
relationship to the $W$-infinity algebras,
which are generated by currents of all integer spins
($s \geq 1$ or $s\geq 2$) including the energy-momentum
tensor and can be viewed as (non-unique) large $N$ limits
of the $W_N$ algebras. There are some very deep connection
between such algebras and the famous KP hierarchy,
which is known to contain all generalized KdV hierarchies,
and the conformal $SL(2,R)/U(1)$ coset model that supports
the black-hole solution in $D=2$ string theory.

The results reported in our previous papers$^{1-4}$ can be
summarized as follows:
1) The centerless linear $W_{1+\infty}$ algebra can be
identified (also see ref. 5) as the first
Hamiltonian structure of the KP hierarchy.
2) There exists a unique non-linear deformation$^2$ of the
$W_{\infty}$ algebra under certain homogeneity assumptions, which we
call $\hat{W}_{\infty}$ and is believed to contain all $W_N$ by
reduction. 3) This $\hat{W}_{\infty}$ algebra is identified as
the (modified) second KP Hamiltonian structure (see also ref. 6).
4) We have found a two boson realization of
the nonlinear $\hat{W}_{\infty}$ algera, in closed form for all
higher spin currents. 5) Using it, we discovered a field
theoretical realization of $\hat{W}_{\infty}$ as a hidden current
algebra in the 2d $SL(2,R)/U(1)$ coset model. 6) The well-known
infinite set of KP Hamiltonians, considered as involutive
$\hat{W}_{\infty}$ charges, generate infinitely many KP flows
in the coset model, all of which preserve the $\hat{W}_{\infty}$
current algerbra.

The above results 3)-6) hold true not only at the classical but
also the {\it quantum} level: 7) A consistent quantum deformation
of the $\hat{W}_{\infty}$ algebra, together with its two boson
realization, is obtained by quantizing the $SL(2,R)_k/U(1)$
coset model. 8) We have been able to construct an infinite set
of commuting quantum $\hat{W}_{\infty}$ charges at level $k$=1.
9) These quantum charges are used to generate a quantum KP hierarchy,
whose Hamiltonian structure is just the quantum $\hat{W}_{\infty}$.

In this talk we give a unified description of these results. Two
long papers containing detailed proofs will be published soon$^{7,8}$.

\vspace{20 pt}
\noindent {\it KP Bi-Hamiltonian Structure, $\hat{W}_{1+\infty}$ and
$\hat{W}_{\infty}$}
\vspace{10 pt}

The integrable KP-Sato hierarchy is an infinite set of nonlinear
partial-differential equations, which reads in the Lax form
\begin{eqnarray}
\frac{\partial L}{\partial t_{m}} = {[(L^{m})_{+},L]}
\end{eqnarray}
where $L$ is a first-order pseudo-differential operator (Psd)
$L = D + \sum^{\infty}_{r=-1}u_{r}D^{-r-1}$, with coefficients
$u_{r}$ functions of $z$ and various times $t_{m} (m=1,2,3, \ldots)$,
and $D\equiv \partial/\partial z$.

The integrability of the KP hierarchy is a direct consequence
of its bi-Hamiltonian structure. To put the KP hierarchy
into the Hamiltonian form
\begin{eqnarray}
\frac{\partial u_{r}}{\partial t_{m}} =
{\{u_{r}(z), \oint H_{m}(z)dz\}},
\end{eqnarray}
one needs to define both the Hamiltonian functions $H_m(z)$ and
the Poisson brackets
\begin{eqnarray}
{\{u_{r}(z),u_{s}(z')\}} = k_{rs}(u_p(z)) \delta (z-z').
\end{eqnarray}

To obtain the bi-Hamiltonian structure, we substitute an arbitrary
Psd of order $N$, $P = \sum_{s=-\infty}^{N} D^{s}p_{s}$,
($N$ can be arbitrarily large) into
\begin{eqnarray}
\hat{K}(P) \equiv  (\hat{L}P)_{+}\hat{L}-\hat{L}(P\hat{L})_{+}=
\sum_{r,s=-1}^{\infty}\hat{k}_{rs}p_{s}D^{-r-1},
\end{eqnarray}
where $\hat{L}=L+c$ (c is a real parameter).

Proposition 1: For fixed pair $(r,s)$, the operator $\hat{k}_{rs}$
becomes stable (i.e. independent of $N$) when $N$ is
sufficiently large. Explicitly, it is given by (for $r,s\geq 0$)
\begin{eqnarray}
\hat{k}_{rs} &=& \sum^{s+1}_{l=0} \left( \begin{array}{c}
s+1 \\ l
\end{array} \right) D^{l}\hat{u}_{r+s+1-l}
-\sum^{r+1}_{l=0} \left( \begin{array}{c}
r+1 \\ l
\end{array} \right) \hat{u}_{r+s+1-l}(-D)^{l} \nonumber\\
& & +\sum^{\infty}_{l=0}{[
\sum^{l+s}_{t=l}(-1)^{l} \left( \begin{array}{c}
t-s-1 \\ l
\end{array} \right)} \nonumber\\
& & {-\sum^{r+s+1}_{t=r+1}\sum^{t-r-1}_{k=0}(-1)^{l+k} \left( \begin{array}{c}
t-k-1 \\ l-k
\end{array} \right) \left( \begin{array}{c}
s \\ k
\end{array} \right) ]}\hat{u}_{t-l-1}D^{l}\hat{u}_{r+s-t}, \nonumber\\
\hat{k}_{r-1} &=& -\sum^{r}_{l=1} \left( \begin{array}{c}
r \\ l
\end{array} \right) \hat{u}_{r-l}(-D)^{l}, \nonumber\\
\hat{k}_{-1s} &=& \sum^{s}_{l=1} \left( \begin{array}{c}
s \\ l
\end{array} \right) D^{l}\hat{u}_{s-l}, ~~~~~ \hat{k}_{-1-1}~=~ -D;
\end{eqnarray}
with $\hat u_{-1} = u_{-1} + c$ , $\hat{u}_r = u_r (r\geq 0)$.
And the operator matrix $k_{rs}$ defines a one-parameter family
of Hamiltonian structures for functions $u_r$.

Proposion 2: The following limits
\begin{eqnarray}
{\{u_{r}(z),u_{s}(z')\}}_{1} &=& \lim_{c\rightarrow \infty}
(\frac{1}{c}\hat{k}_{rs}(z))\delta (z-z')~~~~~r,s\geq 0;\\
{\{u_{r}(z),u_{s}(z')\}}_{2} &=& \lim_{c\rightarrow 0}
\hat{k}_{rs}(z) \delta (z-z')~~~~~~~r,s\geq -1
\end{eqnarray}
coincide, respectively, with the first and second Hamiltonian
structures of the KP hie- rarchy with the Hamiltonian functions
\begin{eqnarray}
H_m^{(1)}={\frac{1}{m+1}} Res L^{m+1},~~~
H_m^{(2)}={\frac{1}{m}} Res L^{m}.
\end{eqnarray}

Note the first KP Hamiltonian structure is linear in $u_r~ (r\geq 0)$:
\begin{eqnarray}
\lim_{c\rightarrow \infty}(\frac{1}{c}\hat{k}_{rs})
= \sum^{s}_{l=0} \left( \begin{array}{c}
s \\ l
\end{array} \right) D^{l}u_{r+s-l}
-\sum^{r}_{l=0} \left( \begin{array}{c}
r \\ l
\end{array} \right) u_{r+s-l}(-D)^{l}.
\end{eqnarray}
(The $u_{-1}$ current becomes decoupled from the algebra.)
Previously we have identified it with the $W_{1+\infty}$
algebra.  We call the nonlinear second Hamiltonian
structure (7) (with $c=0$) as the $\hat{W}_{1+\infty}$ algebra,
in which the current $u_r$ is considered to have spin $r+2$
$(r\geq -1)$. It is a universal $W$-algebra, containing all known
$W$-algebras of the infinite type or of the finite type either
by contraction or by reduction. However, $u_{-1}$ does not evolve
at all: $\partial u_{-1}/\partial t_{m} = 0$. One can always
impose the constraint $u_{-1}=0$ which is second-class in
$\hat{W}_{1+\infty}$.

Proposition 3: With $u_{-1}=0$, the modified second KP Hamiltonian
structure is
\begin{eqnarray}
k_{rs}^{(2D)} &=& \sum^{s+1}_{l=0} \left( \begin{array}{c}
s+1 \\ l
\end{array} \right) D^{l}u_{r+s+1-l} -\sum^{r+1}_{l=0} \left( \begin{array}{c}
r+1 \\ l
\end{array} \right) u_{r+s+1-l}(-D)^{l} \nonumber\\
& & +\sum^{\infty}_{l=0}{[
\sum^{l+s}_{t=l+1}(-1)^{l} \left( \begin{array}{c}
t-s-1 \\ l
\end{array} \right)} \nonumber\\
& & { - \sum^{r+s}_{t=r+1}\sum^{t-r-1}_{k=0}(-1)^{l+k} \left( \begin{array}{c}
t-k-1 \\ l-k
\end{array} \right) \left( \begin{array}{c}
s \\ k
\end{array} \right) ]}u_{t-l-1}D^{l}u_{r+s-t} \nonumber\\
& & -\sum^{r}_{l=1}\sum^{s}_{k=1}(-1)^{l}\left( \begin{array}{c}
r\\l
\end{array} \right) \left( \begin{array}{c}
s\\k
\end{array} \right) u_{r-l}D^{l+k-1}u_{s-k}.
\end{eqnarray}

It can be viewed as the unique nonlinear, centerless deformation of
$W_{\infty}$ under certain natural homogeneity
requirements. We call it as the $\hat{W}_{\infty}$ algebra.

A direct consequence
is the following interesting relation
between $\hat{W}_{\infty}$ and  $W_{1+\infty}$:
\begin{eqnarray}
\oint {\{\hat{W}_{r+1}(z),\hat{W}_{s}(z')\}}_{2D}dz'
= \oint {\{W_{r}(z),W_{s}(z')\}}_{1}dz'.
\end{eqnarray}

\vspace{20 pt}
\noindent {\it Free Field Realization of $\hat{W}_{\infty}$ and
$SL(2,R)/U(1)$}
\vspace{10 pt}

The known realization of $W_{1+\infty}$ with a complex fermion is
simply summarized by
\begin{eqnarray}
L\equiv D+\sum^{\infty}_{r=0}u_{r}D^{-r-1}=D-\bar{\psi}D^{-1}\psi.
\end{eqnarray}
Considering only the spin $s\geq 2$ currents, for the $W_{\infty}$
generators we have
\begin{eqnarray}
(LD)_{-}\equiv\sum^{\infty}_{r=0}v_{r}D^{-r-1}
=-(\bar{\psi}D^{-1}\psi D)_{-}.
\end{eqnarray}
Meanwhile, $W_{\infty}$ can be realized in terms of a complex boson
\begin{eqnarray}
L\equiv D+\sum^{\infty}_{r=0}v_{r}D^{-r-1}=D+\bar{j}D^{-1}j.
\end{eqnarray}
where the bosonic currents $\bar{j}(z)=\bar{\phi}'(z)$,
$j(z)=\phi'(z)$ satisfy
\begin{eqnarray}
{\{\bar{j}(z), j(z')\}} = \partial_{z}\delta(z-z');
{}~~~{\{\bar{j}(z), \bar{j}(z')\}} = {\{j(z), j(z')\}} = 0.
\end{eqnarray}

A nonlinear deformation of this leads to

Proposition 4: The following gives a realization of all
currents $u_r$ of $\hat{W}_{\infty}$:
\begin{eqnarray}
L &\equiv& D+\sum^{\infty}_{r=0}u_{r}D^{-r-1}
= D+\bar{j}\frac{1}{D-(\bar{j}+j)}j \nonumber\\
&=& D+\bar{j}D^{-1}j+\bar{j}D^{-1}(\bar{j}+j)D^{-1}j+ \cdots .
\end{eqnarray}

This realization of $\hat{W}_{\infty}$ naturally emerges in the
$SL(2,R)/U(1)$ coset model. Recall the bosonized
$SL(2,R)_{k}/U(1)$ parafermion currents: (classically we set $k=1$)
\begin{eqnarray}
\psi_{+}=\bar{j}e^{\bar{\phi}+\phi}, ~~~~
\psi_{-}=je^{-\bar{\phi}-\phi}.
\end{eqnarray}
We propose to expand the bilocal product $\psi_{+}(z)\psi_{-}(z')$
in powers of $z-z'$ to all orders.

Proposion 5: The product $\psi_{+}(z)\psi_{-}(z')$ generate all two-boson
$\hat{W}_{\infty}$ currents (16):
\begin{eqnarray}
\bar{j}e^{\bar{\phi}+\phi}(z)je^{-\bar{\phi}-\phi}(z')
= \sum^{\infty}_{r=0}u_{r}(z)\frac{(z-z')^{r}}{r!}.
\end{eqnarray}
Formally this can also be rewritten as
$u_r(z) = \psi_{+}(z)(-\partial_z)^r \psi_{-}(z)$, or
\begin{eqnarray}
L = D + \sum^{\infty}_{r=0}u_{r}D^{-r-1} = D + \psi_{+}D^{-1}\psi_{-}.
\end{eqnarray}

It follows that the Poisson brackets for $\hat{W}_{\infty}$ can be
generated from
\begin{eqnarray}
& & {\{\bar{j}e^{\bar{\phi}+\phi}(z)je^{-\bar{\phi}-\phi}(z-\epsilon),
\bar{j}e^{\bar{\phi}+\phi}(w)je^{-\bar{\phi}-\phi}(w-\sigma)\}} \nonumber\\
&\equiv & {\{\psi_{+}(z)\psi_{-}(z-\epsilon),
\psi_{+}(w)\psi_{-}(w-\sigma)\}}
{}~=~ \sum^{\infty}_{r,s=0} k_{rs}(z) \delta (z-w) \frac{\epsilon^{r}
\sigma^{s}}{r!s!}.
\end{eqnarray}

\vspace{20 pt}
\noindent {\it Involutive KP charges and
$\hat{W}_{\infty}$ Symmetry in $SL(2,R)/U(1)$}
\vspace{10 pt}

The KP charges, $Q_{m}\equiv \oint H_{m}(z) dz$, are known to be
conserved under the KP flows:
\begin{eqnarray}
\frac{\partial}{\partial t_{n}}\oint H_{m}(z,t_{k}) dz =0.
\end{eqnarray}
and they are in involution:
\begin{eqnarray}
{\{Q_{n}^{(2)}, Q_{m}^{(2)}\}}_{2D}
= {\{Q_{n-1}^{(1)}, Q_{m}^{(1)}\}}_{1} =0.
\end{eqnarray}

This implies that in the $SL(2,R)/U(1)$ coset model there exist
infinitely many commuting $\hat{W}_{\infty}$ charges,
which generate an infinite symmetry
\begin{eqnarray}
\frac{\partial u_{r}}{\partial t_{m}} = {\{u_{r}, Q_{m}\}},
{}~~{\rm or}~~
\delta_{m} u_{r} = \epsilon_{m} {\{u_{r}, Q_{m}\}}.
\end{eqnarray}
An important property of these symmetry transformations is that
the $\hat{W}_{\infty}$ current algebra for the $u_r$'s is preserved
under these flows:
\begin{eqnarray}
& & \frac{\partial}{\partial t_{m}}({\{u_{r}(z),u_{s}(z')\}}
-k_{rs}(z)\delta (z-z')) = 0.
\end{eqnarray}
Also these symmetries are abelian to each other,
in the sense that the KP flow in one charge direction is invariant
under the transformation generated by another $\hat{W}_{\infty}$
charge:
\begin{eqnarray}
& & \delta_{m}(\frac{\partial u_{r}}{\partial t_{n}}-{\{u_{r}, Q_{n}\}})
{}~=~ \epsilon_{m}{\{ u_{r}, {\{Q_{n}, Q_{m}\}} \}} ~=~ 0.
\end{eqnarray}
Note that these results hold for the $W_{1+\infty}$ charges as well.

The free boson realization of $\hat{W}_{\infty}$
allows us to construct the KP flows of $j$ and $\bar{j}$:
\begin{eqnarray}
\frac{\partial \bar{j}}{\partial t_{m}} = {\{\bar{j}, Q_{m}\}}, ~~~~
\frac{\partial j}{\partial t_{m}} = {\{j, Q_{m}\}}
\end{eqnarray}
(See also ref. 9.) Note that the Hamiltonian structure of this
hierarchy is simply the fundamental Poisson brackets (15)
for $\bar{j}$ and $j$, which are invariant under
the $\bar{j}$-$j$ flows:
\begin{eqnarray}
\frac{\partial {\{\bar{j},j\}}}{\partial t_{m}} = {\{ {\{\bar{j},Q_{m}\}},
j\}} + {\{\bar{j}, {\{j,Q_{m}\}} \}} = {\{ {\{\bar{j},j\}}, Q_{m} \}} = 0.
\end{eqnarray}

Similarly, the free fermion realization (12) of $W_{1+\infty}$
gives the KP Hamiltonians
\begin{eqnarray}
& & H_{1} = \bar{\psi}\psi, ~~~~ H_{2} = \frac{1}{2}(\bar{\psi}'\psi-
\bar{\psi}\psi'), ~~
 H_{3} = \frac{1}{3}(\bar{\psi}''\psi-\bar{\psi}'\psi'+\bar{\psi}\psi''),
{}~~~{\rm etc.}
\end{eqnarray}
Their charges can also be used to generate a compatible and integrable
$\bar{\psi}$-$\psi$ hierarchy
\begin{eqnarray}
\frac{\partial \bar{\psi}}{\partial t_{m}} = {\{\bar{\psi}, Q_{m}\}}, ~~~~
\frac{\partial \psi}{\partial t_{m}} = {\{\psi, Q_{m}\}}.
\end{eqnarray}


\vspace{20 pt}
\noindent {\it Quantum $\hat{W}_{\infty}$ and Its Two Boson Realization}
\vspace{10 pt}

Because of the connection of the $SL(2,R)/U(1)$ model to the
black holes in 2D string theory, it is necessary to generalize
what we have obtained above to the quantum theory.
We have been succeeded in constructing a quantum version of
the nonlinear $\hat{W}_{\infty}$ algebra, by quantizing the
$SL(2,R)_{k}/U(1)$ coset model, with the bosonized currents
\begin{eqnarray}
\psi_{+}(z;p) &=&
\frac{1}{2}[(1+\sqrt{1-2p})\bar{j}+(1-\sqrt{1-2p})j]
e^{\sqrt{p}(\bar{\phi}+\phi)}, \nonumber\\
\psi_{-}(z;p) &=&
\frac{1}{2}[(1-\sqrt{1-2p})\bar{j}+(1+\sqrt{1-2p})j]
e^{-\sqrt{p}(\bar{\phi}+\phi)}.
\end{eqnarray}
where $\phi(z)$ and $\bar{\phi}(z)$
are a pair of complex bosons,
and their currents $\bar{j}(z)$, $j(z)$ satisfy
\begin{eqnarray}
\bar{j}(z)j(z') \sim \frac{1}{(z-z')^{2}}, ~~~~~
\bar{j}(z)\bar{j}(z') \sim j(z)j(z') \sim 0.
\end{eqnarray}
Note the parameter $p\equiv k^{-1}$ plays the role of
the Planck constant as the essential parameter in quantum
corrections (or quantum deformation).
We have included necessary quantum corrections
in the currents. The classical limit is recovered by
first rescaling $\phi\rightarrow\sqrt{k}\phi$,
$\psi\rightarrow \sqrt{k}\psi$
and then letting $k\rightarrow\infty$.

Generalizing the classical equation (18), we use the whole OPE (up to all
orders),
\begin{eqnarray}
\psi_{+}(z;p)\psi_{-}(z';p)~ = ~
\epsilon^{-2p}\{\epsilon^{-2}+\sum^{\infty}_{r=0}u_{r}(z;p)
\frac{\epsilon^{r}}{r!}\}
\end{eqnarray}
(with $\epsilon\equiv z-z'$) to generate the quantum
$\hat{W}_{\infty}(p)$ currents $u_{r}(z;p)$ in the KP basis. A closed
expression for all $u_{r}(z;p)$, in terms of $\bar{j}$ and $j$ only,
has been derived in ref. 8.

The complete structure of the quantum $\hat{W}_{\infty}(p)$
algebra can be manifested by the OPE's between two currents
$u_{r}(z;p)$ and $u_{s}(w;p)$, as usual in conformal field theory.
In principle, these OPE's can be extracted from the
OPE of four $SL(2,R)_{k}/U(1)$ currents:
\begin{eqnarray}
(\epsilon\sigma)^{2p}(\psi_{+}(z;p)\psi_{-}(z-\epsilon;p))
(\psi_{+}(w;p)\psi_{-}(w-\sigma;p)).
\end{eqnarray}
The closure of the quantum $\hat{W}_{\infty}(p)$ algebra is ensured
by the closure of the OPE's associated with the enveloping algebra
of the $SL(2,R)_{k}$ currents in the neutral sector in
the conformal model. However, once we write the
algebra in the form of OPE's among $u_r$ and $u_s$,
the currents $u_r$ may be considered as independent of each other,
and the associativity of the OPE's automatically leads to closed
Jacobi identities. So our quantum version of the
$\hat{W}_{\infty}$ algebra holds even beyond the
context of the coset model.

The $p=1$ case, corresponding to level $k=1$, may be
interpreted as the ``typical'' quantum case.
For this value of $p$, a lot of expressions simplify.

Proposition 6: In the case of $p=1$, we have
\begin{eqnarray}
u_{0}(z)u_{s}(w) &=& \frac{-1}{z-w} u_{s}'(w) + {\it ~terms~in~other~
powers~of~}(z-w)^{-1}, \nonumber\\
u_{1}(z)u_{s}(w) &=& \frac{-2}{z-w} [\sum^{s}_{l=1}(-1)^{l}
\left( \begin{array}{c}
s\\l
\end{array} \right) (u_{0}^{(l)}u_{s-l}) +\frac{u_{s+1}'}{(s+1)}
+\frac{(-1)^{s}u_{0}^{(s+2)}}{(s+1)(s+2)}](w) \nonumber\\
& & +{\it ~terms~in~other~powers~of~}(z-w)^{-1}.
\end{eqnarray}

\newpage
\noindent {\it Commuting Quantum $\hat{W}_{\infty}$ Charges
and Quantum KP Hierarchy}
\vspace{10 pt}

The above OPE's provide sufficient
information for constructing an infinite set of commuting
quantum $\hat{W}_{\infty}$ charges, whose densities are quantum
deformation of the classical Hamiltonian functions (8).
For generic $p$, it is not hard to construct
\begin{eqnarray}
H_{2}(z;p) &=& \frac{1}{(1-2p)}u_{0}(z;p),
{}~~~H_{3}(z;p) ~=~ u_{1}(z;p)+\frac{1}{2}u_{0}'(z;p), \nonumber\\
H_{4}(z;p) &=& u_{2}(z;p)+u_{1}'(z;p)+\frac{(5+4p)}{15}u_{0}''(z;p)
+\frac{p}{(1-2p)}(u_{0}u_{0})(z;p).
\end{eqnarray}
In the ``typical'' quantum case with $p=1$, we can exploit
the two general OPE (34) to explicitly construct an infinite number
of commuting quantum $\hat{W}_{\infty}$ charges $Q_{m}$.

Let us recall that in conformal field theory the local
product $(AB)(z)$ is defined by
\begin{eqnarray}
(AB)(z) = \oint_{z} \frac{A(w)B(z)}{w-z} dw
\end{eqnarray}
in which the small contour of integration encircles $z$. This
product is neither commutative nor associative. The symmetric local
product of $N$ local operators is defined to be the totally
symmetrized sum of their multiple local products taken from the left:
\begin{eqnarray}
\langle A_{1}A_{2}\cdots A_{N} \rangle = \frac{1}{N!}\sum_{P\{i\}}
(\cdots ((A_{i_{1}}A_{i_{2}})A_{i_{3}})\cdots A_{i_{N}})
\end{eqnarray}
where $P\{i\}$ denotes the summation over all possible permutations.

To construct quantum $\hat{W}_{\infty}$ charges,
let us assign a degree 1 to $\partial_z$ and $r+2$ to $u_r$, and
assume that the quantum charge-density $H_{m}(z)$
is homogeneous and of degree $m$, and led by the highest-spin
current $u_{m-2}$ with unity coefficient.
We require their charges commute with each other:
\begin{eqnarray}
{[\oint_{0} H_{n}(z)dz , \oint_{0} H_{m}(w)dw]} = 0.
\end{eqnarray}

Proposition 7:  The commutativity with $Q_2$ is always satisfied:
\begin{eqnarray}
{[\oint_{0} H_{2}(z)dz , \oint_{0} H_{m}(w)dw]} = 0
\end{eqnarray}

Thus the first set of nontrivial equations in eq.(38) start with $n=3$.
Here $H_{3}$ can only be $u_{1}$ plus a derivative of $u_{0}$;
the latter does not contribute to the charge $Q_3$. We have shown$^{8}$
that charges $Q_{m}$ are completely determined
by the requirement of commutativity with $Q_3$ alone. Both amusingly
and amazingly, the so-determined charges $Q_{m}$ automatically
commute with each other. (A similar situation has happened before for
commuting charges for the quantum KdV equation.$^{10}$)

Proposition 8: The charges
\begin{eqnarray}
Q_{m} = \oint_{0} \sum_{l} \sum_{\{i\}}C_{i_{1}i_{2}\cdots i_{l}}(m)
\langle u_{i_{1}}u_{i_{2}}\cdots u_{i_{l}} \rangle (z)dz
\end{eqnarray}
with the elegant expression for the coefficients
\begin{eqnarray}
C_{i_{1}i_{2}\cdots i_{l}}(m) = \frac{(-1)^{l-1}(l-1)!(m-2)!}{d_{1}!d_{2}!
\cdots d_{k}!i_{1}!i_{2}!\cdots i_{l}!},
\end{eqnarray}
provides a unique solution to eq.(38) for arbitrary $m$.
Here for given number of currents in the product, $l$,
the summation is over all partitions $\{i_{k}\}$ of $m-2l$,
satisfying $i_{1}+i_{2}+\cdots +i_{l}=m-2l$; and here $d$'s denote
the degeneracies in the partition:
$0\leq i_{1}=i_{2}=\cdots =i_{d_{1}}$ $<i_{d_{1}+1}=\cdots
=i_{d_{1}+d_{2}}$ $<\cdots =i_{d_{1}+d_{2}+\cdots +d_{k}(=l)}$.

We emphasize that in the proof of this Proposition, we have used only
the OPE's in eq.(34); the two boson realization (32) of the
$\hat{W}_{\infty}$ currents $u_{r}$ is never used.

Now it is straightforward to construct a quantum version of
the KP hierarchy in the Hamiltonian form (2).
We use the quantum $\hat{W}_{\infty}$ as the Hamiltonian structure,
and the densities of the quantum charges $Q_{m}$ as corresponding
Hamiltonian functions $H_{m}$. The latter naturally generate
an infinite set of compatible flows in times $t_{m}$ $(m=1,2,\ldots)$:
\begin{eqnarray}
\frac{\partial u_{r}}{\partial t_{m}} = {[u_{r}, Q_{m+1}]}.
\end{eqnarray}
The mutual commutativity
of these quantum charges implies that they are
conserved charges of the flows (42):
\begin{eqnarray}
\frac{\partial Q_{n}}{\partial t_{m}} = {[Q_{n}, Q_{m+1}]} = 0.
\end{eqnarray}
This ensures the complete integrability of the hierarchy (42).
We call it the $p=1$ quantum KP hierarchy and view it as a
desired quantum version of the classical KP hierarchy.
To justify this, we have computed the first three quantum KP charges
for generic $p$ and have shown that they coincide with the charges
(40) at $p=1$ and reduce to the classical KP charges when
$p\rightarrow 0$. Also we have caluculated the first few
quantum KP flows with generic $p$:
\begin{eqnarray}
& & \frac{\partial u_{0}}{\partial t_{1}} = (1-2p)u_{0}',
{}~~ \frac{\partial u_{1}}{\partial t_{1}} = (1-2p)u_{1}',
{}~~ \frac{\partial u_{0}}{\partial t_{2}} = (1-2p)(2u_{1}'+u_{0}''),
\nonumber\\
& & \frac{\partial u_{1}}{\partial t_{2}} = (2-3p)u_{2}'+(1-p)u_{1}''
+\frac{(1+2p)p}{6}u_{0}''' +p(u_{0}u_{0})', \nonumber\\
& & \frac{\partial u_{0}}{\partial t_{3}} = 3(1-2p)(u_{2}'+u_{1}'')
+\frac{(3-4p-p^{2})}{3}u_{0}''' +3p(u_{0}u_{0})'.
\end{eqnarray}
Manipulating the last three equations, we obtain the first
dynamically nontrivial quantum evolution equation for the
current $u_{0}(z;p)$:
\begin{eqnarray}
6(2-3p)\frac{\partial u_{0}'}{\partial t_{3}}
-9\frac{\partial^{2} u_{0}}{\partial t_{2}^{2}} =
(3-p-16p^{2}+18p^{3})u_{0}'''' +18(1-p)p(u_{0}u_{0})''.
\end{eqnarray}
Rescaling $u_0\rightarrow u_0/p$ and taking $p\rightarrow 0$, this
reduces to the classical KP equation, as claimed.

It should also be possible to obtain an alternative
quantum deformation of the classical KP hierarchy
through deforming its much simpler first Hamiltonian
structure -- the $W_{1+\infty}$. Because of its
linearity, quantization of $W_{1+\infty}$ should be straightforward.
In the KP basis, the complete structure of
the quantum $W_{1+\infty}$ is neatly manifested by the following OPE's:
\begin{eqnarray}
u_{r}(z)u_{s}(w) &=& \sum^{r}_{l=0}\frac{r!}{(r-l)!}
\frac{u_{r+s-l}(z)}{(z-w)^{l+1}} - \sum^{s}_{l=0}(-1)^{l}\frac{s!}{(s-l)!}
\frac{u_{r+s-l}(w)}{(z-w)^{l+1}} \nonumber\\
& & + \frac{(-1)^{s}cr!s!}{(z-w)^{r+s+2}} + O(z-w).
\end{eqnarray}
The remaining issue is to construct an complete set of infinitely many
commuting quantum charges in accordance to (46), in order for the
associated quantum KP hierarchy to be integrable.

To conculde, we remark that it would be interesting to clarify
the role of our world-sheet nonlinear $\hat{W}_{\infty}$ symmetry
in physics of 2D black holes, and to invent a formalism for 2D
strings in terms of the KP hierarchy or $\hat{W}_{\infty}$ constraints.

\vspace {20 pt}
\noindent {\it Acknowledgement}: The work was supported in part by
US NSF grant PHY-9008452.
\vspace{10 pt}

\newpage
{\it REFERENCES}
\begin{itemize}

\item[1.] F. Yu and Y.-S. Wu, Phys. Lett. B263 (1991) 220.
\item[2.] F. Yu and Y.-S. Wu, Nucl. Phys. B373 (1992) 713.
\item[3.] F. Yu and Y.-S. Wu, Phys. Rev. Lett. 68 (1992) 2996.
\item[4.] F. Yu and Y.-S. Wu, Utah preprint UU-HEP-92/11, May 1992;
to be published in Phys. Lett. B.
\item[5.] K. Yamagishi, Phys. Lett. B259 (1991) 436.
\item[6.] J. M. Figueroa-O'Farrill, J. Mas and E. Ramos, Phys. Lett.
B266 (1991) 298; preprint BONN-HE-92/20, US-FT-92/7 or KUL-TF-92/20.
\item[7.] F. Yu and Y.-S. Wu, Utah preprint UU-HEP-92/07, August 1992.
\item[8.] F. Yu and Y.-S. Wu, Utah preprint UU-HEP-92/12, August 1992.
\item[9.] D. A. Depireux, Laval preprint LAVAL-PHY-21-92;
J. M. Figueroa-O'Farrill, J. Mas and E. Ramos,
preprint BONN-HE-92/17, US-FT-92/4 or KUL-TF-92/26.
\item[10.] R. Sasaki and I. Yamanaka, Commun. Math. Phys. 108 (1987) 691;
Adv. Stud. in Pure Math. 16 (1988) 271.
\end{itemize}

\end{document}